\documentclass[twocolumn,showpacs]{revtex4}

\usepackage{graphicx}%
\usepackage{dcolumn}
\usepackage{amsmath}

\makeatletter
\def\btt#1{\texttt{\@backslashchar#1}}%
\DeclareRobustCommand\bblash{\btt{\@backslashchar}}%
\makeatother

\newcommand{\DIR}{.}

\begin{document}

\preprint{physics/0506189}

\title[Short Title]{Basics of Modelling the Pedestrian Flow}

\author{Armin Seyfried}
\email{A.Seyfried@fz-juelich.de}
\affiliation{
Central Institute for Applied Mathematics, 
Research Centre J\"ulich,
52425 J\"ulich,
Germany
}

\author{Bernhard Steffen}
\email{B.Steffen@fz-juelich.de}
\affiliation{
Central Institute for Applied Mathematics, 
Research Centre J\"ulich,
52425 J\"ulich,
Germany
}

\author{Thomas Lippert}
\email{Th.Lippert@fz-juelich.de}
\affiliation{
Central Institute for Applied Mathematics, 
Research Centre J\"ulich,
52425 J\"ulich,
Germany
}

\date{\today}
 
\begin{abstract}
For the modelling of pedestrian dynamics we treat persons as self-driven 
objects moving in a continuous space. On the basis of a modified social force 
model we qualitatively analyze the influence of various approaches for the 
interaction between the pedestrians on the resulting velocity-density relation. 
To focus on the role of the required space and remote force we choose a 
one-dimensional model for this investigation. For those densities, where in two 
dimensions also passing is no longer possible and the mean value of the velocity 
depends primarily on the interaction, we obtain the following result: If the 
model increases the required space of a person with increasing current velocity, 
the reproduction of the typical form of the fundamental diagram is possible. 
Furthermore we demonstrate the influence of the remote force on the 
velocity-density relation.
\end{abstract}


\maketitle
\section{Introduction}
\label{intro}

Microscopic models are state of the art for computer simulation of pedestrian 
dynamics. The modelling of the individual movement of pedestrians results in a 
description of macroscopic pedestrian flow and allows e.g. the evaluation of 
escape routes, the design of pedestrian facilities and the study of more theoretical 
questions. For a first overview see \cite{PED01,PED03}. The corresponding models can 
be classified in two categories: the cellular automata models 
\cite{MURA99,BLUE00,NAGA02,SCHA01,SCHR02} and models in a continuous space 
\cite{HELB95,HOOG02,THOM95,KOEN01}. We focus on models continuous in space. They 
differ substantially with respect to the `interaction' between the pedestrians and 
thus to the update algorithms as well. The social force model for example assumes, 
among other things, a repulsive force with remote action between the pedestrians 
\cite{HELB95,MOLN96,HELB00a,HELB00b,HELB01,HELB01b,HELB03}. Other models treat 
pedestrians by implementing a minimum inter-person distance, which can be 
interpreted as the radius of a hard body \cite{THOM95,KOEN01}. 

One primary test, whether the model is appropriate for a quantitative description 
of pedestrian flow, is the comparison with the empirical velocity-density relation 
\cite{SCHR02b,SCHA04,HOOG02b,RIMEA}. In this context the fundamental 
diagram of Weidmann \cite{WEID93} is frequently cited. It describes the 
velocity-density relation for the movement in a plane without bottlenecks, 
stairs or ramps. A multitude of causes can be considered which determine this 
dependency, for instance friction forces, the 'zipper' effect \cite{HOOG05}  
and marching in step \cite{NAV69,SEYF05}. As shown in \cite{SEYF05} the empirical 
velocity-density relation for the single-file movement is similar to the relation 
for the movement in a plane in shape and magnitude. This surprising conformance indicates, 
that lateral interferences do not influence the fundamental diagram at least up to a 
density-value of $4.5 \,m^{-2}$. This result suggests that it is sufficient to 
investigate the pedestrian flow of a one-dimensional system without loosing the 
essential macroscopic characteristics. We modify systematically the social force 
model to achieve a satisfying agreement with the empirical velocity-density relation 
(fundamental diagram). Furthermore we introduce different approaches for the 
interaction between the pedestrians to investigate the influence of the required 
space and the remote action to the fundamental diagram.


\section{Modification of the Social Force Model}

\subsection{Motivation}

The social force model was introduced by \cite{HELB95}. It models the 
one-dimensional movement of a pedestrian $i$ at position $x_i(t)$ with
velocity $v_i(t)$ and mass $m_i$ by the equation of motions
 
\begin{equation}
\frac{d x_i}{d t} = v_i \quad \quad m_i \frac{d v_i}{d t} = F_i = \sum_{j\neq i} 
F_{ij}(x_j,x_i,v_i). 
\end{equation} 

The summation over $j$ accounts for the interaction with other pedestrians. We assume 
that friction at the boundaries and random fluctuations can be neglected and thus 
the forces are reducible to a driving and a repulsive term $F_i=F^{drv}_i+F^{rep}_i$. 
According to the social force model \cite{HELB95} we choose the driving term 

\begin{equation}
F^{drv}_{i}= m_i \frac{v^0_i-v_i}{\tau_i},
\end{equation}

where $v^0_i$ is the intended speed and $\tau_i$ controls the acceleration. 
In the original model the introduction of the repulsive force 
$F^{rep}_i$ between the pedestrians is motivated by the observation that 
pedestrians stay away from each other by psychological reasons, e.g. to secure 
the private sphere of each pedestrian \cite{HELB95}. The complete model reproduces 
many self-organization phenomena like e.g. the formation of lanes in bi-directional 
streams and the oscillations at bottlenecks 
\cite{HELB95,MOLN96,HELB00a,HELB00b,HELB01,HELB01b,HELB03}. In the publications 
cited, the exact form of this repulsive interaction changes and the authors 
note that most phenomena are insensitive to its exact form \cite{HELB01b}. 
We choose the force as in \cite{HELB00a}. 

\begin{equation}
F^{rep}_{i}=\sum_{j\neq i} - \nabla A_i \left(\|x_j-x_i\|-d_{i}\right)^{-B_i}
\label{FREP}
\end{equation}

The hard core, $d_i$, reflects the size of the pedestrian $i$ acting 
with a remote force on other pedestrians. Without other constraints a repulsive 
force which is symmetric in space can lead to velocities which are in opposite direction 
to the intended speed. Furthermore, 
it is possible that the velocity of a pedestrian can exceed the intended speed 
through the impact of the forces of other pedestrians. In a two-dimensional 
system this effect can be avoided through the introduction of additional forces 
like a lateral friction, together with an appropriate choice of the interaction 
parameters. In a one-dimensional system, where lateral interferences are excluded, 
a loophole is the direct limitation of the velocities to a certain interval 
\cite{HELB95,MOLN96}.

Another important aspect in this context is the dependency between the current 
velocity and the space requirement. As suggested by Pauls in the 
extended ellipse model \cite{PAULS04} the area taken up by a pedestrian increase 
with increasing speed. Thompson also based his model on the assumption, that 
the velocity is a function of the inter-person distance \cite{THOM95}. Furthermore 
Schreckenberg and Schadschneider observed in \cite{SCHR02b,SCHA04}, that in 
cellular automata model's the consideration, that a pedestrian occupies all cells 
passed in one time-step, has a large impact on the velocity-density relation. 
Helbing and Moln\'ar note in \cite{HELB95} that the range of the repulsive interaction 
is related to step-length. Following the above suggestion we specify 
the relation between required space and velocity for a one-dimensional system.
In a one-dimensional system the required space changes to a required length $d$.
In \cite{SEYF05} it was shown that for the single-file movement the relation 
between the required lengths for one pedestrian to move with velocity $v$ 
and $v$ itself is linear at least for velocities $0.1\,m/s < v < 1.0\,m/s$. 

\begin{equation}
 d =  a + b\,v \quad \mbox{with} \quad a=0.36\,m \quad \mbox{and} \quad b=1.06\,s 
\end{equation}

Hence it is possible to determine one fundamental microscopic parameter, $d$, of 
the interaction on the basis of empirical results. This allows focusing on the 
question if the interaction and the equation of motion result in a correct 
description of the individual movement of pedestrians and the impact of 
the remote action. Summing up, for the modelling of regular motions of pedestrians 
we modify the reduced one-dimensional social force model in order to meet the following
properties: the force is always pointing in the direction of the intended velocity $v_i^0$; 
the movement of a pedestrian is only influenced by effects which are directly positioned 
in front; the required length $d$ of a pedestrian to move with velocity $v$ 
is $d=a + b \, v$. 

\subsection{Interactions}
\label{INTER}

To investigate the influence of the remote action both a force which treats 
pedestrians as simple hard bodies and a force according to Equation \ref{FREP}, 
where a remote action is present, will be introduced. For simplicity we set 
$v^0_i > 0$, $x_{i+1} > x_i$ and the mass of a pedestrian to $m_i=1$.\\

{\bf{Hard bodies without remote action}}

\begin{eqnarray}
 F_i(t) = \left\{\begin{array}{r@{\quad:\quad}l}
              \frac{v^0_i-v_i(t)}{\tau_i} & x_{i+1}(t)-x_i(t) > d_{i}(t) \\
                     - \delta(t) v_i(t) & x_{i+1}(t)-x_i(t) \leq d_{i}(t) 
\end{array}\right.
\label{NODIST}
\end{eqnarray}
with
\begin{equation}
d_i(t) = a_i + b_i v_i(t)\nonumber 
\end{equation}

The force which acts on pedestrian $i$ depends only on the position, 
its velocity, and the position of the pedestrian $i+1$ in front. 
As long as the distance between the pedestrians is larger than the required 
length, $d_i$, the movement of a pedestrian is only influenced by the driving 
term. If the required length at a given current velocity is larger 
than the distance the pedestrian stops (i. e. the velocity becomes zero). 
This ensures that the velocity of a pedestrian is restricted to the interval 
$v_i = [0,v_i^0]$ and that the movement is only influenced by the pedestrian 
in front. The definition of $d_i$ is such that the required length 
increases with growing velocity.\\

{\bf{Hard bodies with remote action}}

\begin{equation}
F_i(t) =  \left\{ \begin{array}{r@{\quad:\quad}l}
             G_i(t)  & v_i(t) > 0 \\
             \max\left(0,G_i(t)\right) & v_i(t) \leq 0 
\end{array} \right.
\label{DIST}
\end{equation}
with
\begin{eqnarray}
\quad G_i(t)= \frac{v^0_i-v_i(t)}{\tau_i} 
             -e_{i}\left(\frac{1}{x_{i+1}(t)-x_i(t)-d_{i}(t)}\right)^{f_i}\nonumber
\end{eqnarray}
and
\begin{equation}
d_i(t) = a_i + b_i v_i(t)\nonumber
\end{equation}

Again the force is only influenced by actions in front 
of the pedestrian. By means of the required length, $d_i$, the range of the interaction is 
a function of the velocity $v_i$. Two additional parameters, $e_i$ and $f_i$, have to
be introduced to fix the range and the strength of the force. Due to the remote action 
one has to change the condition for setting the velocity to zero. The above definition 
assures that the pedestrian $i$ stops if the force would lead to a negative velocity. 
With the proper choice of $e_i$ and $f_i$ and sufficiently small time steps this 
condition gets active mainly during the relaxation phase. Without remote action this 
becomes important. The pedestrian can proceed when the influence of the driving term 
is large enough to get positive velocities.\\

This different formulation of the forces requires different update algorithms, which 
will be introduced in the next section. A special problem stems from the periodic 
boundary conditions enforced for the tests of the fundamental diagram, as these 
destroy the ordering by causality, which otherwise could avoid blocking situations.

\subsection{Time stepping algorithm}

The social force model gives a fairly large system of second order ordinary differential 
equations. For the hard body model with remote action, where the right hand side 
of the ODE's is continuous along the solution, an explicit Euler method with a time 
step of $\Delta t=0.001\,s$ was tested and found sufficient. Within that time, the 
distance between two persons does not change enough to make the explicit scheme 
inaccurate. 

The situation for the hard body model without remote force is more complicated. 
Here the right hand side is a distribution, and the position of the Dirac spikes is 
not known a priory. Hence the perfect treatment is an adaptive procedure, where each 
global time step is restricted to the interval up to the next contact. Unfortunately, 
this is a complicated and time consuming process. For a simple time step we choose 
the following procedure: Each person is advanced one step ($\Delta t=0.001\,s$) 
according to the local forces. If after this step the distance to the person in front 
is smaller than the required length, the velocity is set to zero and the position to 
the old position. Additionally, the step of the next following person is reexamined. 
If it is still possible, the update is completed. Otherwise, again the velocity is 
set to zero and the position is set to the old position, and so on. This is an 
approximation to the exact parallel update. It is not completely correct, however. 
To test its independence from the ordering of persons, computations using different 
orders were performed. The differences were minute and not more than expected from 
reordering of arithmetic operations.  


\section{Results}

To enable a comparison with the empirical fundamental diagram of the single-file 
movement \cite{SEYF05} we choose a system with periodic boundary conditions and a 
length of $L=17.3\,m$. For both interactions we proofed that for system-sizes of 
$L=17.3, 20.0, 50.0\,m$ finite size effects have no notable influence on the results. 
The values for the intended speed $v^0_i$ are distributed according to a 
normal-distribution with a mean value of $\mu=1.24\,m/s$ and $\sigma=0.05\,m/s $. 
In a one-dimensional system the influence of the pedestrian with the smallest 
intended speed masks jamming effects which are not determined by individual properties. 
Thus we choose a $\sigma$ which is smaller than the empirical value and 
verified with $\sigma=0.05, 0.1, 0.2\,m/s$, that a greater 
variation has no influence to the mean velocities at larger densities. 

In reality the parameters $\tau,a,b,e$ and $f$ are different for every pedestrian $i$ 
and correlated with the individual intended speed. But we know from experiment 
\cite{SEYF05} that the movement of pedestrians is influenced by phenomena like 
marching in step and in reality the action of a pedestrian depends on the 
entire situation in front and not only on the distance to the next person. 
Therefore it's no point to attempt to give fully accurate values of this 
parameter and we may choose identical values for all pedestrians. We tested 
variations of the parameters and found that the behavior changes continuously. 
According to \cite{HELB03}, $\tau=0.61\,s$ is a reliable value. 

For every run we set at $t=0$ all velocities to zero and distribute 
the persons randomly with a minimal distance of $a$ in the system. After $3 \times 10^5$ 
relaxation-steps we perform $3 \times 10^5$ measurements-steps. At every step we 
determine the mean value of the velocity over all particles and calculate the mean 
value over time. The following figures present the dependency 
between mean velocity and density for different approaches to the interaction 
introduced in section \ref{INTER}. To demonstrate the influence of a required length 
dependent on velocity we choose different values for the parameter $b$. 
With $b=0$ one get simple hard bodies.

\begin{figure}[ht]
\begin{center}
\includegraphics[width=1.15\columnwidth]{\DIR/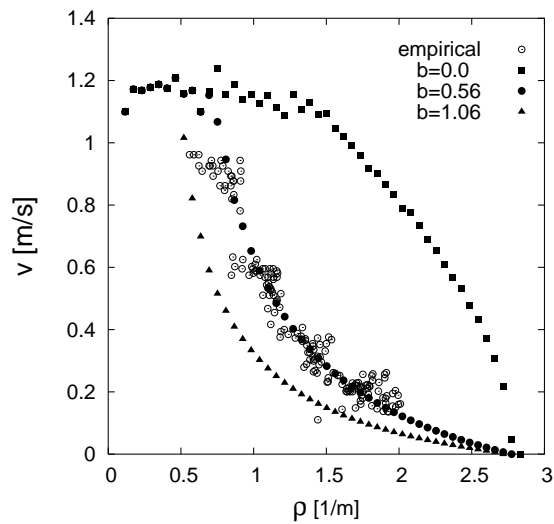}
\end{center}
\caption[]{Velocity-density relation for hard bodies with $a=0.36\,m$ and without 
a remote action in comparison with empirical data from \cite{SEYF05}. The filled 
squares result from simple hard bodies. The introduction of a required length 
with $b=0.56\,s$ leads to a good agreement with the empirical data.}
\label{HDSPH}
\end{figure}

Figure \ref{HDSPH} shows the relation between the mean values of walking speed and
density for hard bodies with $a=0.36\,m$ and without remote action, according to 
the interaction introduced in Equation \ref{NODIST}. If the required length is 
independent of the velocity, one gets a negative curvature of the function 
$v=v(\rho)$. The velocity-dependence controls the curvature and $b=0.56\,s$ results 
in a good agreement with the empirical data. With $b=1.06\,s$ we found a difference 
between the velocity-density relation predicted by the model and the empirical 
fundamental diagram. The reason for this discrepancy is that the interaction and 
equation of motion do not describe the individual movement of pedestrian correctly. 
To illustrate the influence of the remote force, we fix the parameter 
$a=0.36\,m,\,\,b=0.56\,s$ and set the values which determine the remote force to 
$e=0.07\,N$ and $f=2$.

\begin{figure}[ht]
\begin{center}
\includegraphics[width=1.15\columnwidth]{\DIR/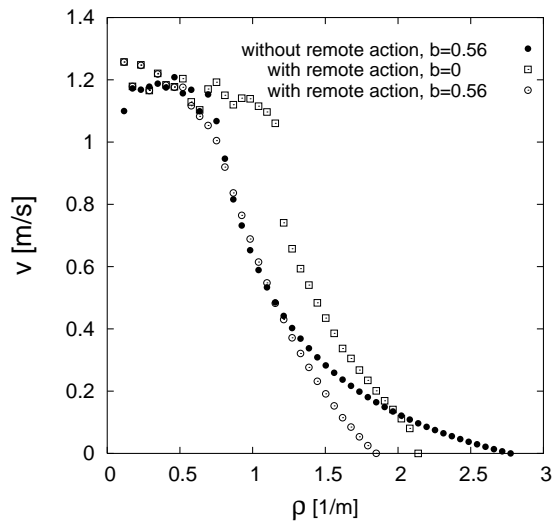}
\end{center}
\caption[]{Velocity-density relation for hard bodies with remote action in 
comparison with hard bodies without a remote action (filled circles). Again 
we choose $a=0.36\,m$ and $b=0.56\,s$. The parameter $e=0.07\,N$ and $f=2$ 
determine the remote force. With $b=0$ one gets a qualitative different fundamental 
diagram and a gap for the resulting velocities.}
\label{SZFRC}
\end{figure}

\begin{figure}[ht]
\begin{center}
\includegraphics[width=1.2\columnwidth]{\DIR/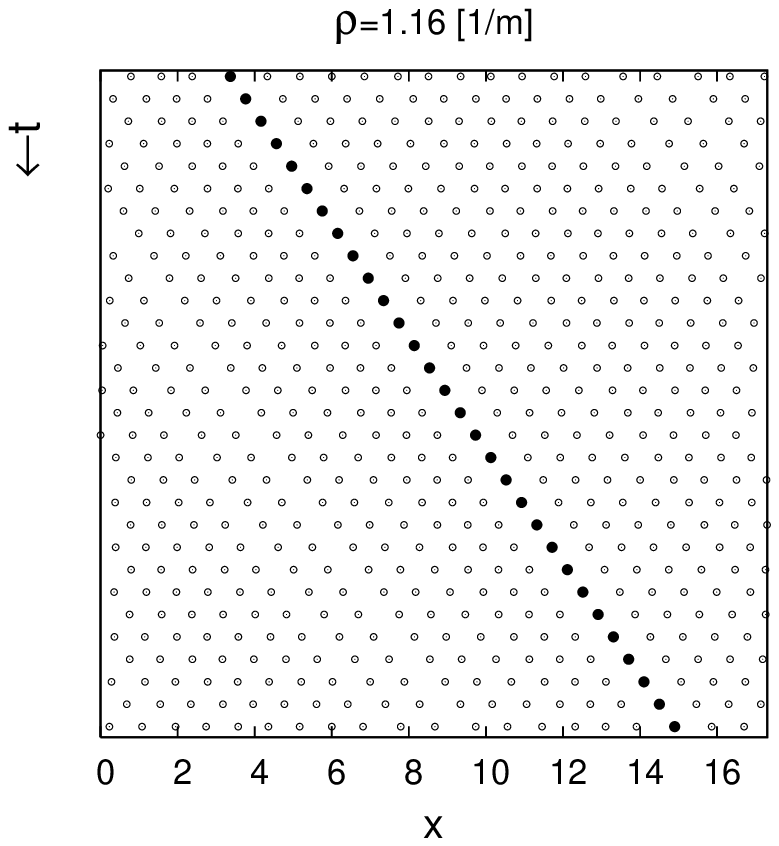}
\includegraphics[width=1.2\columnwidth]{\DIR/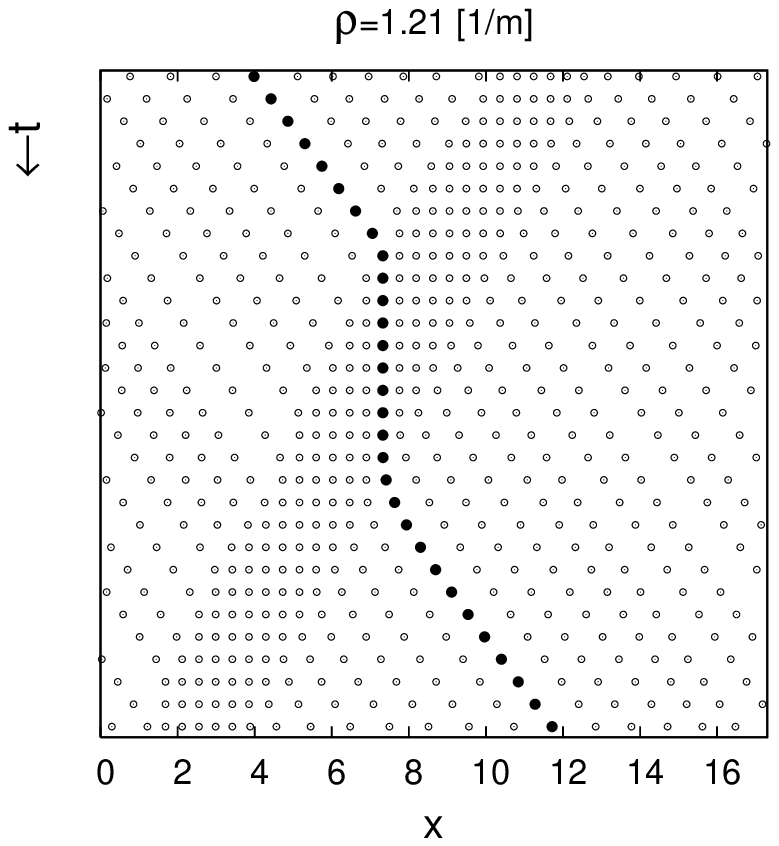}
\end{center}
\caption[]{Time-development of the positions for densities near the velocity-gap, see 
Figure \ref{SZFRC}. For $\rho>1.2\,m^{-1}$ density waves are observable. Some 
individuals leave much larger than average gaps in front.}
\label{DNSW}
\end{figure}

The fundamental diagram for the interaction with remote action according to 
Equation \ref{DIST} is presented in Figure \ref{SZFRC}. The influence is small 
if one considers the velocity-dependence of the required length. But with $b=0$ 
one gets a qualitative different fundamental diagram. The increase of the velocity 
can be expected due to the effective reduction of the required length. The gap at 
$\rho\approx1.2\,m^{-1}$ is surprising. It is generated through the development of 
distinct density waves, see Figure \ref{DNSW}, as are well known from highways.
From experimental view we have so far no hints to the development of strong density 
waves for pedestrians \cite{SEYF05}. The width of the gap can be changed by 
variation of the parameter $f$ which controls the range of the remote force. Near 
the gap the occurrence of the density waves depends on the distribution of the 
individual velocities, too.

\section{Discussion and summary}

For the investigation of the influence of the required space and remote action  
on the fundamental diagram we have introduced a modified one-dimensional social force 
model. The modifications warrant that in the direction of intended speed negative
velocities do not occur and that the motion of the pedestrians is influenced
by objects and actions directly in front only. If one further takes into account that 
the required length for moving with a certain velocity is a function of the current 
velocity the model-parameter can be adjusted to yield a good agreement with the 
empirical fundamental diagram. This holds for hard bodies with and without remote 
action. The remote action has a sizeable influence on the resulting velocity-density 
relation only if the required length is independent of the velocity. In this case 
one observes distinct density waves, which lead to a velocity gap in the fundamental 
diagram.

Thus we showed that the modified model is able to reproduce the empirical 
fundamental diagram of pedestrian movement for a one-dimensional system, if it
considers the velocity-dependence of the required length. For the model parameter 
$b$ which correlates the required length with the current velocity, we have found 
that without remote action the value $b=0.56\,s$ results in a velocity-density relation 
which is in a good agreement with the empirical fundamental diagram. However, from 
the same empirical fundamental diagram one determines $b=1.06\,s$, see \cite{SEYF05}. 
We conclude that a model which reproduces the right macroscopic dependency between density and 
velocity does not necessarily describe correctly the microscopic situation, and 
the space requirement of a person at average speed is much less than the average space 
requirement. This discrepancy may be explained by the 'short-sightedness' 
of the model. Actually, pedestrians adapt their speed not only to the person 
immediately in front, but to the situation further ahead, too. This gives a much 
smoother movement than the model predicts.

The above considerations refer to the simplest system in equilibrium and with 
periodic boundary conditions. In a real life scenario like a building evacuation, 
where one is interested in estimates of the time needed for the clearance 
of a building and the development of the densities in front of bottlenecks, one 
is confronted with open boundaries and conditions far from equilibrium. We assume 
that a consistency on a microscopic level needs to be achieved before one can accurately
describe real life scenarios. The investigation presented provides a basis for a 
careful extension of the modified social force model and an upgrade to two dimensions 
including further interactions.

\begin{acknowledgments}
We thank Oliver Passon for careful reading and Wolfram Klingsch for discussions.
\end{acknowledgments}


\end{document}